\documentclass[10pt,a4paper]{article}
\usepackage{graphpap,epsfig,amssymb,graphicx,textcomp}
\usepackage[utf8]{inputenc}
\usepackage[english]{babel}
\begin{document}
\begin{center}{\Large {\bf Spin flip statistics and spin wave interference}\\
{\bf patterns in Ising Ferromagnetic Films: A Monte Carlo study}}\end{center}

\vskip 1cm

\begin{center}{\it Muktish Acharyya}\\
{\it Department of Physics, Presidency University}\\
{\it 86/1 College Street, Calcutta-700073, INDIA}\\
{E-mail:muktish.physics@presiuniv.ac.in}\end{center}

\vskip 1cm

\noindent {\bf Abstract:} The spin wave interference is studied in two
dimensional Ising ferromagnet driven by two coherent spherical magnetic field
waves by Monte Carlo simulation. The spin waves are found to 
propagate and interfere according to the
classic rule of interference pattern generated by two point sources. The 
interference pattern of spin wave is observed in one boundary of the lattice.
The interference pattern is detected and studied by spin flip statistics at
high and low temperatures. The destructive interference is manifested as the
large number of spin flips and vice versa.  

\newpage

\noindent {\bf I. Introduction:}

The spin wave excitation in the ferromagnets is a fundamental excitation 
which is well studied\cite{chaikin} in anisotropic
Heisenberg ferromagnets. Several experimental studies were perform to observe the 
behaviours\cite{prl} of spin waves and the interference pattern of two coherent spin waves. 
The eigenmodes of spinwaves of square permaloy dots are studied\cite{gubbiotti} by
Brillouin ligh scattering. The Brillouin scattering of lights by spin waves in
ferromagnetic nanorods are studied\cite{stashkevich}. To study the propagation of
spin waves across thin magnetic film samples and experimental scheme has been proposed
\cite{kruglyak} which is based on the creation of picosecond pulses of strongly localized
effective magnetic field via ultrafast optical irradiation of a spatially deposited
exchange bias. 
The method of realising spin wave logic gate is investigated 
experimentally\cite{sch}.

The experiments on the interference of spin waves are also perfomed recently.
A few of those may be mentioned here.
The modulation of propagating spin wave 
amplitude in ${\rm Ni}_{81}{\rm Fe}_{19}({\rm Py})$
films, resulting from constructive and destructive interference of spin waves has
been studied\cite{sankha}. The spin wave interference from rising and falling edges of 
electrical pulses has also been studied\cite{yang} to see the effect of the electrical
pulses width of input excitation on the generated spin waves in a 
${\rm Ni}_{81}{\rm Fe}_{19}({\rm Py})$ strip using
pulse inductive time domain measurements. Current induced localized spin wave interference in a
ferromagnetic nanowire with a domain wall is investigated\cite{yoon}
by simulation.

Several theoretical investigations are also done in past few years. The nonequilibrium
behaviours of interference of spin waves, in the system of two dimensional electron gas with
both Rashba and Dresselhaus (001) spin-orbit coupling, are investigated\cite{chen} 
by Green's function
within tight binding framework. Spin wave interference in microscopic ferromagnetic rings
were studied and series of quantized modes in the vortex state were found\cite{pod}.

The discussions made above reveals that the interference of spin waves in
ferromagnetic materials becomes an important field of
modern research. If the dynamical modes of spin waves are to be understood clearly,
one has to study the time dependent response of spin in ferromagnetic models. The
nonequilibrium response of ferromagnet driven by time dependent magnetic field would
be helpful for this. {\it What will be the right approach to study the spin wave excitation
and its interference, in model ferromagnets, from statistical mechanical point of view ?
What will be the statistics of spin flips if the spin wave travels through the ferromagnetic
sample and how does it reflected in the interference pattern of spin waves ?} 
These questions 
are not yet addressed in the literature (as far as the knowledge of this author is concerned)
and this is the main motivation of this paper. Before going to the core of the problem,
it would be convenient for the reader if a brief introduction is given. First
of all, in the model ferromagnet, like Ising model, one has to know how
the spin bands travel, if it is driven by propagating magnetic field wave.
In the next para, some recent studies, on the propagating 
wavelike spin bands in the Ising ferromagnet driven by propagating magnetic
field wave, will be mentioned.

The nonequilibrium responses of model ferromagnets, 
driven by oscillating magnetic field is studied in details\cite{rev,marev}.  
The hysteresis and dynamic phase transition are two
major responses giving rise to many nonequilibrium phenomena. However, the field
oscillates in time and remains uniform over the space (at any particular instant)
in all those above mentioned stdies. If the field has both spatial and temporal
variations, it would lead to further interesting facts. It may be visualised as the
propagating magnetic field wave with a well defined frequency and wavelength. If
such wave passes through the ferromagnet, the coherent motion of the spin clusters
are observed. These magnetic wave may generally be of two types, namely, plane wave
and spherical wave. 

The propagation of plane spin wave is obevserved \cite{wikipedia,youtube} 
and dynamic phase transition is found\cite{phys}
in Ising ferromagnetic film driven by plane propagating magnetic wave.
Recently the propagation of spin clusters and existence of various
dynamic phases are observed \cite{rfim} in random field Ising ferromagnet. In normal
ferromagnet, the propagation of circular spin waves are observed
\cite{spw} in Ising ferromagnet
and multiple nonequilibrium phase transitions are found.
The importance of all these studies mentioned above are 
not merely pedagogical. Recent experiments \cite{expt}
on Permalloy (50 nm thick) excited by ultrashort (150 fs) 
laser pulse gives rise 
to the propagation of circular spin waves.

The question naturally arises in mind, what will happen if two coherent spin waves
interfere ? Can one study the interference pattern in this case using the tools of
statistical mechanics? Being motivated by
this idea the high field spin wave interference pattern,
in driven Ising ferromagnetic film,  is studied and reported in this
paper. The paper is organised as follows: the Model and the Monte Carlo simulation
method are described in section II, the numerical results are presented in section III, the
letter ends with a summary in section IV.
\vskip 1cm

\noindent {\bf II. Model:}

The Hamiltonian of a two dimensional
 Ising ferromagnet driven by magnetic field wave is:

\begin{equation}
H(t) = -J \sum'_{x,y,x',y'} s(x,y,t)s(x',y',t) - \sum_{x,y} h(x,y,t)s(x,y,t)
\end{equation}

\noindent Where, $s(x,y,t) = \pm 1$ is Ising spin at position (x,y) and in 
time $t$. $J(>0)$ is the ferromagnetic spin-spin interaction strength. 
The first term represents nearest neighbour spin spin interaction. The
prime over the summation indicates the sum over distinct nearest neighbour
pairs.
In this model,
the time dependent Hamiltonian $H(t)$ 
(equation-(1))
is measured in the unit of $J$.
$h(x,y,t)$ is the magnetic field (measured in the unit of $J$)
at position (x,y) in time $t$ coming from
a spherically propagating magnetic wave represented as:

\begin{equation}
h(x,y,t) = {{h_0 {\rm cos} 2\pi (ft - y/{\lambda})} \over r}
\end{equation}

\noindent The amplitude, frequency and the wavelength are represented by
$h_0$, $f$ and $\lambda$ respectively. 
$r$ is the distance from the source. If the source is placed at the origin
$r=\sqrt{x^2 + y^2}$.
The model is defined on a square lattice of linear size $L$.
Two coherent point sources of spherically propagating magnetic wave 
(represented in equation-(2)) are  
placed on the x-axis. The periodic boundary conditions are applied in both
directions.

\newpage

\noindent {\bf III. Methodology:}

In the above mentioned model, the initial spin configuration is chosen
as statistically and randomly chosen 50 percent spins take the value +1.
This is a very high temperature configuration. Two coherent point sources 
of spherical magnetic waves are symmetrically placed in the middle of
one end of the lattice with a separation $\delta$. In this simulation,
one source is placed at a distance $\delta/2$ in the right side from the center
and the other one is placed in the left side of the center. The Monte Carlo
simulation is employed here with single spin flip Metropolis dynamics. The
probability of spin flip is:

\begin{equation}
P(s(x,y,t) \to -s(x,y,t)) = {\rm Min}[1, {\rm exp}({{-\Delta E} \over {kT}})]
\end{equation}

\noindent where $\Delta E$ is the change in energy due to spin flip, $k$ is
Boltzmann constant and $T$ is the temperature of the system. Here, the 
temperature $T$ is measured in the unit of $J/k$.
A spin is chosen randomly and flipped with the probability mention above
in equation-3.
$L^2$ such random move defines one Monte Carlo Step per Spin (MCSS) and defines
the time unit in the problem.
The steady state
configuration is acheived after $2\times10^4$ MCSS and 
it is checked that this number is adequate. 

In the opposite (to that where the sources are placed) side of the lattice
the number of flips of spins at each site is calculated.

\vskip 1cm

\noindent {\bf IV. Results:}

Here, linear size of the lattice, $L=100$ is considered. Two coherent
point sources of spherical magnetic waves are placed symmetrically about
the central site of one end (say bottom line). The separation 
$\delta$ is taken equal to 19 lattice unit. This choice is arbitrary, just
to make it of the order of the wavelength of the sources. The wavelength
($\lambda$)
of the source is taken 5 (lattice unit). The frequency $f=0.01$
and the amplitude of each source is taken $h_0=2000$. As a result
one complete cycle of the propagating magnetic wave requires 100 MCSS.
The choice of this value of frequency $f=0.01$ requires 100 MCSS. One can
visualize the changes of the wave pattern (see fig-1) within 25 MCSS which
can be obtained in a few minutes in $L=100$ in intel core i5 processor.
The value of the amplitude of the field $h_0=2000$ is chosen quite
large to see the 
significant effect of interference on the other side of the lattice.
It may be noted that, the strength of field falls as ${{1} \over {r}}$
as mentioned in eqn-2.

In figure-1 the propagation of spin waves are shown by plotting the
spin configurations for three different time instants. The 
propagation spin waves is very similar to that of water waves for two 
coherent sources\cite{wikipedia}. At the opposite end the interference pattern
is observed. The propagating magnetic field will interfere and the position
of destructive interference will be reflected in the number of flips 
(averaged over the full cycle of the propagating magnetic field) of the
spins in that region. In this region the cooperatively interacting term
will govern the probability of spin flip and the number of spin flip will
be more. On the other hand, the square average of the 
superposed value of propagating magnetic field due to two sources, is very high
in the region of constructive interference. 
In this region, the spins will follow the directions of the 
magnetic fields.
As a result, the number of flips
of the spin will be less. The number of flips of the spins are shown 
in figure-2, as
a functions of position (in the top line of the lattice). Here, the 
positions of the minima of
the square averaged magnetic field are same as the positions of the average number of flips of
the Ising spins. These are studied for three different temperatures. As
the temperature decreases the number of flips decreases
(as the probability of flip decreases due to Metropolis rule
described in eqn-3), however the 
steady interference pattern of the spin waves are maintained.
 
\vskip 1cm 

\noindent {\bf V. Discussion:}

In this letter, the interference pattern of spin waves in Ising ferromagnet
due to two coherent sources is studied by Monte Carlo simulation. The steady
interference pattern of spin waves is recognized as the average number of flips of
the spins. The destructive interference of magnetic wave gives rise to large
averaged number of spin flips as the dynamics is governed by the cooperative term in
the Hamiltonian. On the other hand, the constructive interference of 
superposed value of magnetic waves keeps the spin along the direction of the field and hence
reduce the number of flips of the spins. This steady interference pattern 
of spin waves persists over a wide range of temperature. It may be mentined 
here that the circular ripples of spin waves are experimentally detected
in ferromagnetic sample by ultrashort laser pulse\cite{expt}. It would be
interesting to study the interference of propagating field 
driven spin waves experimentally. 

The interference of spin waves in Ising ferromagnetic film studied in this
paper is an appeal to the experimentalists to see the effect in real 
ferromagnetic samples irradiated by strong optical sources which may lead
to interesting and technologically important features in the field of
spintronics\cite{bader} 
\vskip 1cm

\noindent {\bf Acknowledgements:} The library facilities provided by Calcutta
University is gratefully acknowledged.

\newpage
%%%%%%%%%%%%%%%%%%%%%%%%%%%%%%%%%%%%%%%%%%%%%%%%%%%%%%%%%%%%%%%%%%%%%%%%%%%%%%%%
\begin{figure}[h]
\begin{center}
\begin{tabular}{c}
\resizebox{9cm}{!}{\includegraphics[angle=0]{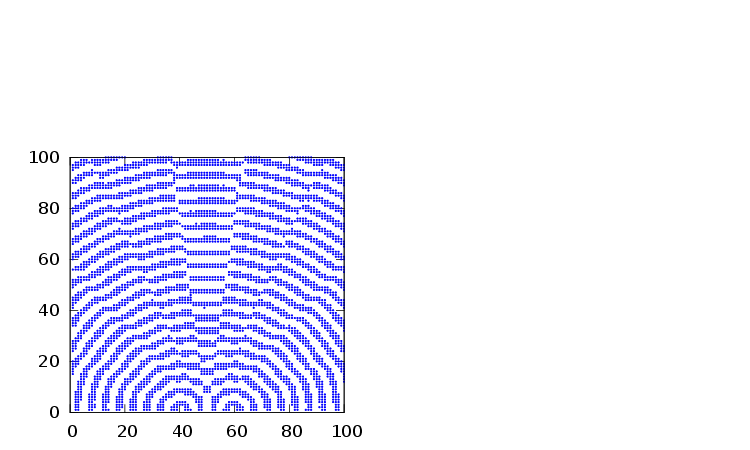}}
\\
\resizebox{9cm}{!}{\includegraphics[angle=0]{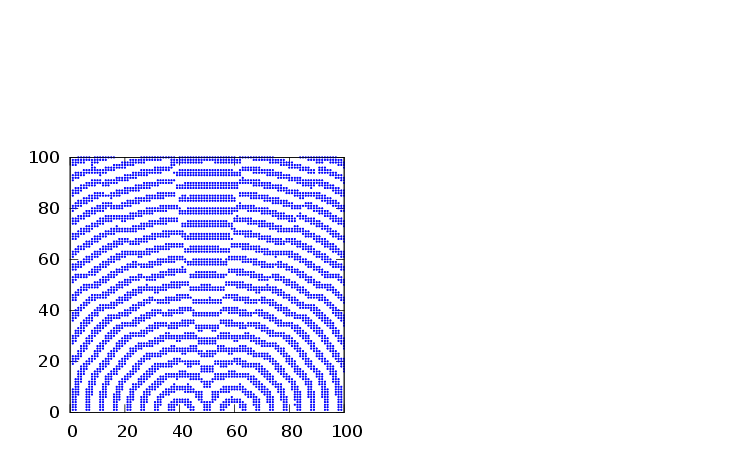}}
\\
\resizebox{9cm}{!}{\includegraphics[angle=0]{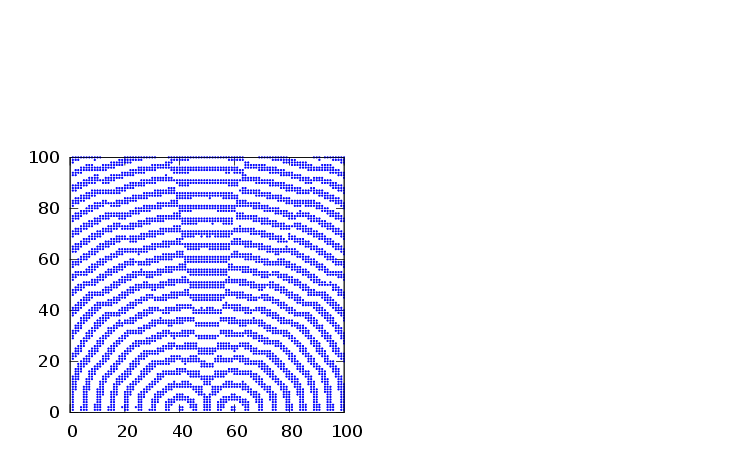}}
          \end{tabular}
\caption
{Propagation of spin waves due to two coherent sources
of spherical magnetic field wave. Different figures represent snapshots in 
different times. Top one is taken at $t = 1950$, middle
one is taken at $t=1975$ and the bottom one is taken at $t=2000$.
Here, $L=100$, $h_0=2000$, $f=0.01$, $\lambda=5$ and $T=2.30$.}
\label{fig:pattern}

\end{center}
\end{figure}
%%%%%%%%%%%%%%%%%%%%%%%%%%%%%%%%%%%%%%%%%%%%%%%%%%%%%%%%%%%%%%%%%%%%%%%%%%%%%%%%%%

\newpage
%%%%%%%%%%%%%%%%%%%%%%%%%%%%%%%%%%%%%%%%%%%%%%%%%%%%%%%%%%%%%%%%%%%%%%%%%%%%%%%%%%
\begin{figure}[h]
\begin{center}
\begin{tabular}{c}
\resizebox{9cm}{!}{\includegraphics[angle=0]{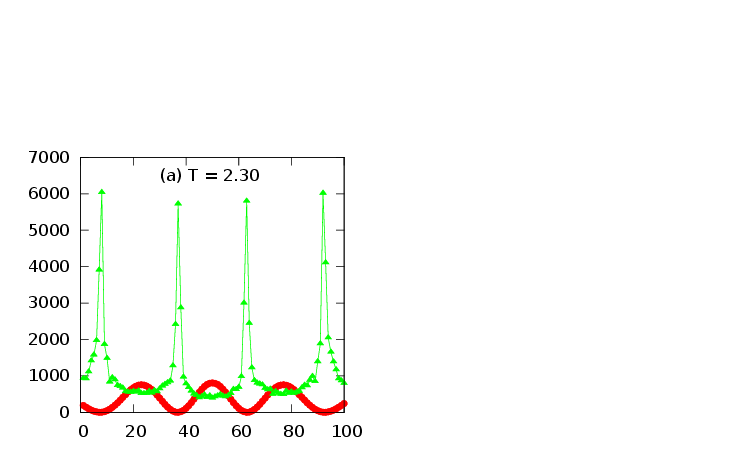}}
\\
\resizebox{9cm}{!}{\includegraphics[angle=0]{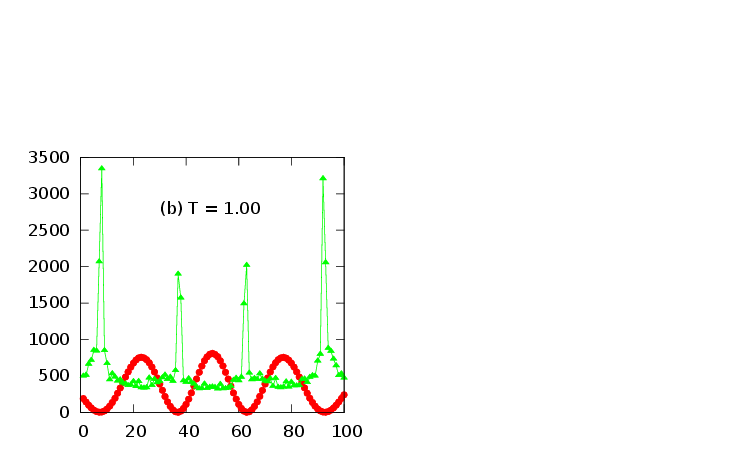}}
\\
\resizebox{9cm}{!}{\includegraphics[angle=0]{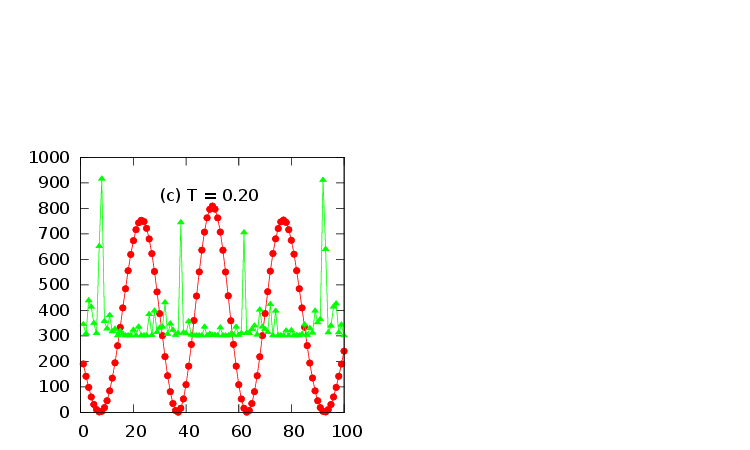}}
          \end{tabular}
\caption
{The spin wave intereference pattern. The number of flips of the
spins are plotted against the position (along x-axis) on the top line of the lattice.
The average number of flips of the spins is represented by (Green triangles) and the square average
(over the full cycle) of the superposed propagating field is shown by (Red bullets). The patterns
at different temperatures,
(a) at $T=2.30$, (b) at $T=1.00$ and (c) at $T=0.20$.
Here, $L=100$, $h_0=2000$, $f=0.01$ and $\lambda=5$.}
\label{fig:flip}

\end{center}
\end{figure}
%%%%%%%%%%%%%%%%%%%%%%%%%%%%%%%%%%%%%%%%%%%%%%%%%%%%%%%%%%%%%%%%%%%%%%%%%%%%%%%%%%

\end{document}